# Propagation-invariant strongly longitudinally polarized toroidal pulses

Ren Wang[1,2*], Ding-Tao Yang[1], Tao Xin[1], Shuai Shi[1], Bing-Zhong Wang[1], and Yijie Shen[3,4*]

[1] Institute of Applied Physics, University of Electronic Science and Technology of China, Chengdu 611731, China

[2] Yangtze Delta Region Institute (Huzhou), University of Electronic Science and Technology of China, Huzhou 313098, China

[3] Division of Physics and Applied Physics, School of Physical and Mathematical Sciences, Nanyang Technological University, Singapore 637378, Singapore

[4] Centre for Disruptive Photonic Technologies, The Photonics Institute, Nanyang Technological University, Singapore 637378, Singapore

*Correspondence to: Ren Wang (rwang@uestc.edu.cn) and Yijie Shen (yijie.shen@ntu.edu.sg)

## Abstract

Recent advancements in optical, terahertz, and microwave systems have unveiled non-transverse optical toroidal pulses characterized by skyrmionic topologies, fractal-like singularities, space-time nonseparability, and anapole-exciting ability. Despite this, the longitudinally polarized fields of canonical toroidal pulses notably lag behind their transverse counterparts in magnitude. Interestingly, although mushroom-cloud-like toroidal vortices with strong longitudinal fields are common in nature, they remain unexplored in the realm of electromagnetics. Here, we present strongly longitudinally polarized toroidal pulses (SLPTPs) which boast a longitudinal component amplitude exceeding that of the transverse component by over tenfold. This unique polarization property endows SLPTPs with robust propagation characteristics, showcasing nondiffracting behavior. The propagation-invariant strongly longitudinally polarized field holds promise for pioneering light-matter interactions, far-field superresolution microscopy, and high-capacity wireless communication utilizing three polarizations.

## Introduction

The longitudinal optical component, parallel to the direction of propagation, has given rise to a diverse array of intriguing structured lights, including optical knots [1]-[4], skyrmions [5]-[10], hopfions [11]-[14], möbius strips [15]-[19], and so on. The longitudinal component endows the optical field with complex and stable topologies, resulting in topological quasiparticles of light [20]. These non-transverse quasiparticles possess significant potential applications in diverse fields such as light–matter interactions [21]-[27], microscopy [28]-[31], metrology [32]-[34], and communications [35]-[39].

As a captivating propagatable non-transverse optical toroidal vortices, toroidal pulses were initially conceived by Hellwarth and Nouchi, drawing upon Ziolkowski's rigorous space-time solutions within the framework of Maxwell's equations [40]-[42]. These pulses, resembling toroidal bubbles (Fig. 1(a1)), exhibit skyrmionic topologies [6] and demonstrate space-time inseparability [43]-[44], thereby potentially inducing electromagnetic anapoles [45]. Recent progress has seen the generation of toroidal bubble-type pulses (Fig.

1(a2)) across various domains, including optics [46], terahertz [47], and microwave [48] systems. In fact, the toroidal vortex phenomena in nature encompass an alternative form exemplified by the mushroom cloud, as depicted in Fig. 1(b1). Mushroom-cloud-type toroidal pulses (Fig. 1(b2)) are anticipated to exhibit pronounced longitudinal polarization field components, yet they remain largely unexplored.

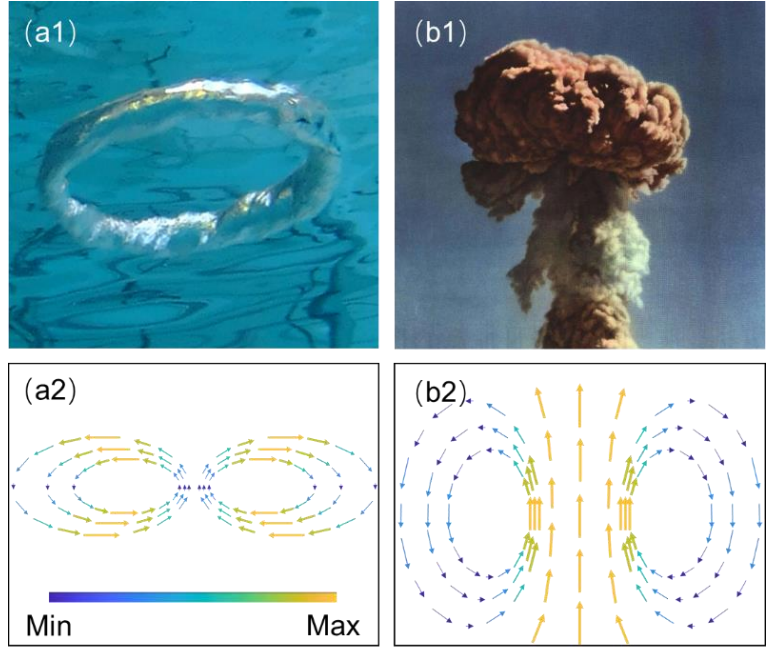

**Fig. 1: Two kinds of toroidal vortex phenomena.** (a1) Toroidal bubble underwater, adapted from [Wikipedia](Wikipedia). (b1) Toroidal mushroom cloud, adapted from [Wikipedia](Wikipedia). Schematics on longitudinal sections of (a2) elementary toroidal pulse, resembling toroidal bubble, and (b2) strongly longitudinally polarized toroidal pulses, resembling the mushroom cloud with toroidal vortex topology.

In this paper, we delve into the longitudinal and transverse components of a more generalized toroidal pulse and introduce propagation-invariant strongly longitudinally polarized toroidal pulses (SLPTP) that endure the singular polarization upon propagation. Notably, the amplitude of longitudinally polarized component exceeds that of the transverse component by over tenfold.

## Results

In 2004, Lekner proposed the so-called oscillatory wavefunction by using $f(s) = q_2 q_1 e^{-k_{oc.}s} / (s + q_2)$ for the general solution obtained by Hillion [49], $\psi(\rho, z, t) = f(s) / [q_1 + i(z - ct)]$, of scalar Helmholtz's wave equation, where $k_{oc.}$ is a positive oscillatory wavenumber and $s = \rho^2 / q_1 + i(z - ct) - i(z + ct)$ [50], where the parameters $q_2$ and $q_1$ represent respectively the depth of the focal region and effective wavelength when compared to a Gaussian beam.. The Ziolkowski's generating function for the widely researched toroidal pulses is apparently an elementary case of Lekner's generating function corresponding to $k_{oc.} = 0$ [40-48]. The solutions with non-zero $k_{oc.}$ have never been studied before. In this work, we explore the polarization characteristic of toroidal pulses with non-zero $k_{oc.}$ and introduce a SLPTP that endure the singular polarization upon propagation. Please see the supplementary material for the detailed derivation of the electromagnetic fields of the SLPTP.

The absolute amplitude distributions of the electric field of an elementary toroidal pulse with $k_{oc.}=0$ and a SLPTP with $k_{oc.}=1.463$ $q_1$ at various times are shown in Fig. 2. For both of the pulses, $q_2=500q_1$. Based on Fig. 2, it is evident that, during propagation, the maxima of the elementary toroidal pulse remain consistently situated on both sides of the propagation axis ($x$=0), whereas the maxima of the SLPTP are consistently located on the propagation axis. A visual inspection of the electric field polarization vectors reveals that the primary electric field maxima of the SLPTP and the elementary toroidal pulse correspond predominantly to longitudinal and transverse polarizations, respectively. To quantitatively illustrate the relative magnitudes of different polarization vectors within the toroidal pulses, we superimposed all vectors pointing in the same direction in Figs. 2(a1~a3 and b1~b3) and plotted the magnitudes of these vector sums in Fig. 2c. Fig. 2c demonstrates that the electric field orientations of the elementary toroidal pulse are primarily concentrated in the transverse direction, whereas the sum of longitudinal electric field components in the SLPTP significantly exceeds other components. This indicates that the elementary toroidal pulse with $k_{oc.}=0$ and the SLPTP with $k_{oc.}=1.463$ $q_1$ can be characterized as strongly transverse and strongly longitudinal pulses, respectively. Furthermore, Fig. 2 illustrates that both the SLPTP and the elementary toroidal pulse maintain their respective polarization characteristics during propagation.

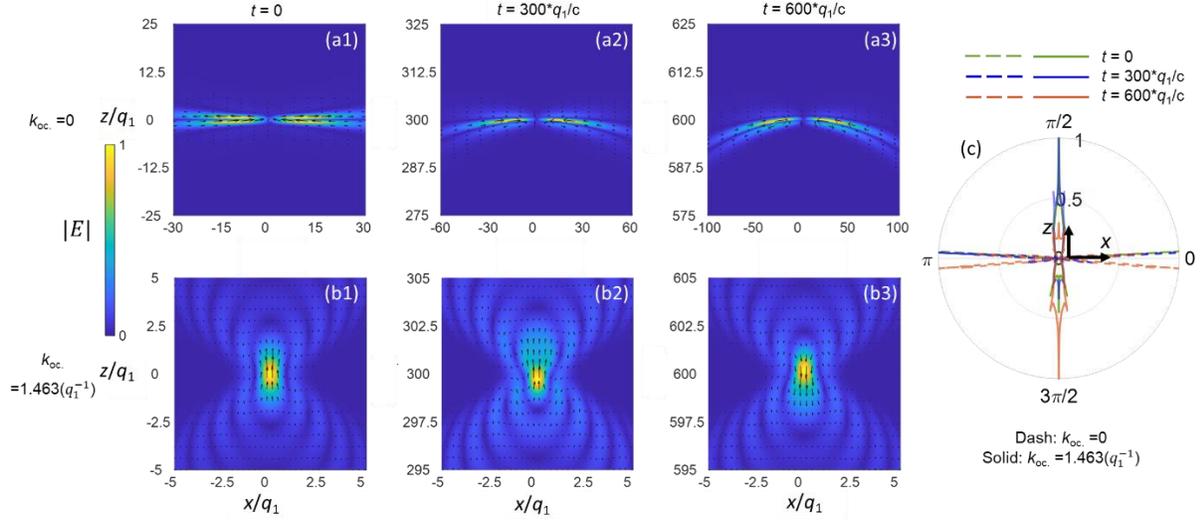

**Fig. 2: Electric field vector of elementary toroidal pulses and SLPTPs.** The absolute amplitude distributions of the electric field on x-z plane of (a1~a3) an elementary toroidal pulse with $k_{oc.}$=0 and (b1~b3) a SLPTP with $k_{oc.}$=1.463 $q_1$ at various times (a1 and b1) $t = 0$, (a2 and b2) $t = 300 \; q_1/c$, and (a3 and b3) $t = 600 \; q_1/c$. The black arrows in (a1~a3) and (b1~b3) indicate the polarization of electric field at different positions. (c) The absolute amplitude of the vector sum for the electric fields in (a1~a3) and (b1~b3) with each direction. For both of the pulses, $q_2$=500$q_1$.

The oscillatory wavenumber, $k_{oc.}$, plays a crucial role in determining the longitudinal polarization characteristic of the SLPTP, as evident from Fig. 3(a). It is observed that as $k_{oc.}$ decreases below 1.463/$q_1$, the ratio of the maximum values of the longitudinal component, $|E_z|$, to the transverse component, $|E_\rho|$, gradually increases. When $k_{oc.}$ approaches zero, the transverse component significantly exceeds the longitudinal component, exhibiting a strongly transverse polarization characteristic. Conversely, as $k_{oc.}$ nears 1.463/$q_1$, the longitudinal component dominates over the transverse component, displaying a strongly longitudinal polarization characteristic. Notably, when $k_{oc.}$ falls within the range of 1/$q_1$ ~2/$q_1$, the longitudinal component exceeds the transverse component by a factor of 8 or more, reaching an elevenfold difference at $k_{oc.}$ =1.463/$q_1$. In contrast, the parameter $q_2$ has little influence on the longitudinal polarization characteristic. Fig. 3(b) illustrates the ratio of the maximum values of $|E_z|$ and $|E_\rho|$ when the center of the toroidal pulses is positioned at different $z$-values. It is apparent that the ratio $|E_z|_{max}/|E_\rho|_{max}$ exhibits a periodic variation when $k_{oc.} = 1.463/q_1$. This periodicity arises from the asynchronous periodic changes in the transverse and longitudinal components during propagation, resulting in the maxima of these components not occurring simultaneously. Detailed insights into this phenomenon are provided in the supplementary materials. However, as evident from Fig. 3(b), when $k_{oc.} = 1.463/q_1$, the longitudinal component of the toroidal pulses generally dominates over the transverse component throughout the propagation process. This indicates that the toroidal pulses can maintain their strongly longitudinal polarization characteristic during propagation.

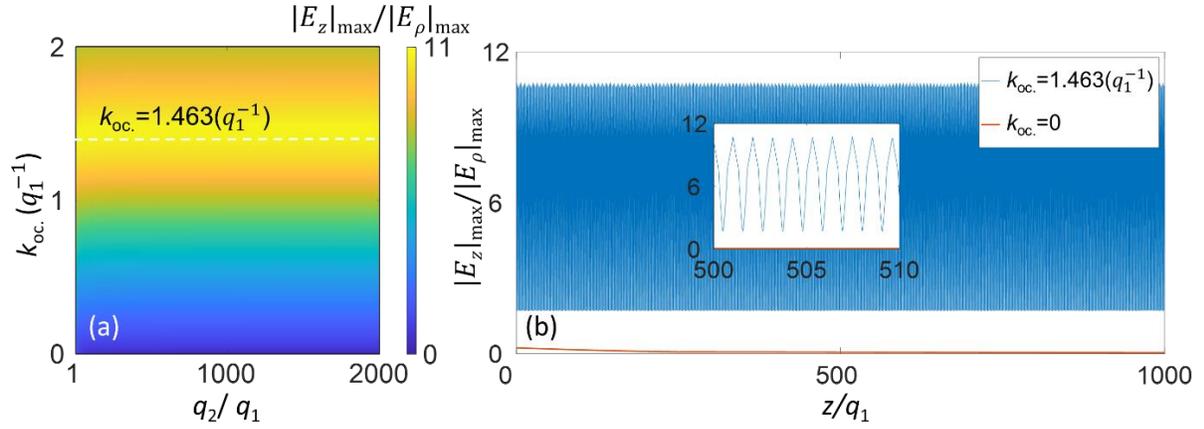

**Fig. 3: The ratio of the maximum values of longitudinal and transverse electric components $|E_z|_{max}$ and $|E_\rho|_{max}$ of the toroidal pulses with different oscillatory wavenumber $k_{oc.}$.** (a) Versus $q_2/q_1$ and (b) versus the pulse center position at different $z$-values. For (a), z=0; for (b), $q_2$=500$q_1$.

The propagation-invariant characteristics can be analyzed using light cone diagrams [51],[52]. Fig. 4 illustrates the plane wave spectra of both longitudinally and transversely polarized components of the SLPTP. It is evident from the figure that on the $\omega$/c - $k_z$ plane, the plane wave spectra approximate straight lines, forming a 45° angle with the $k_z$ axis. This alignment corresponds to the spatial spectral characteristics of elliptical diffraction-free beams propagating at the speed of light [51]. The invariant propagation can also been intuitively observed from the spectral traces and frequency spectra at different position, further information can be found in the supplementary materials. Consequently, the SLPTP emerges as a new propagation-invariant light sheet with strongly longitudinal polarization.

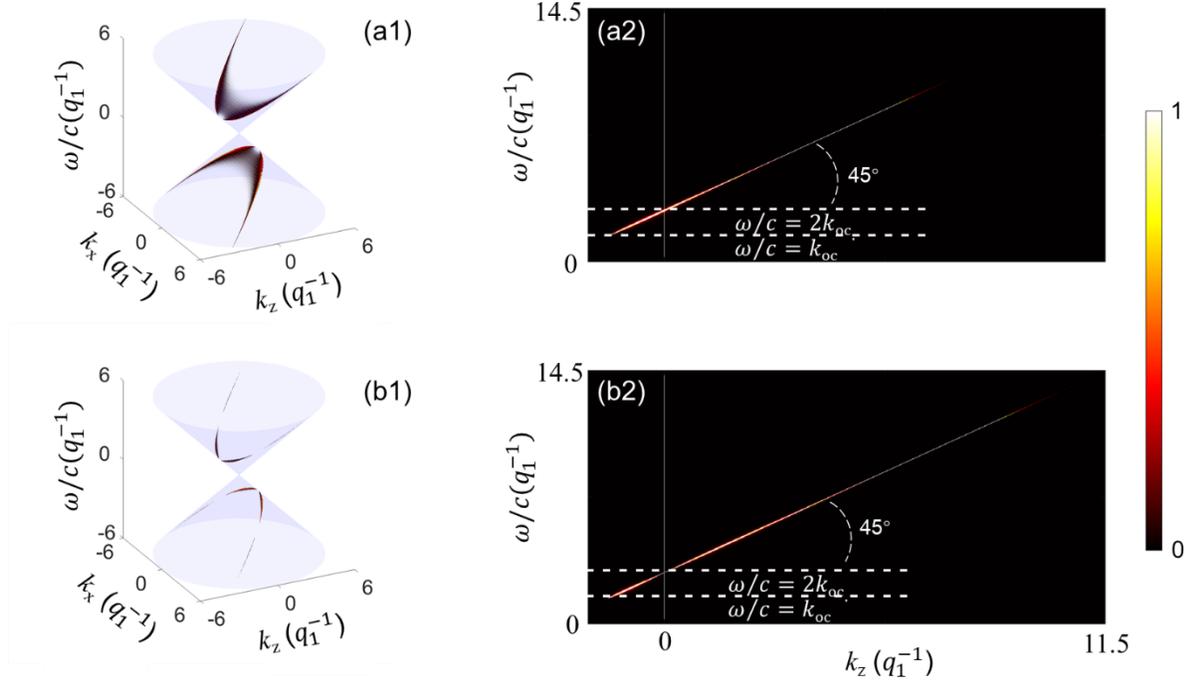

**Fig. 4: Plane wave spectra of the (a1 and a2) longitudinally and (b1 and b2) transversely polarized components of the SLPTP with $k_{oc.}$=1.463 $q_1$.** The light cone is represented by the lilac surface in (a1 and b1). The corresponding projections of longitudinally and transversely polarized components into the $k_z$-$\omega$/c plane are shown in (a2) and (b2), respectively.

From Fig. 4, it is also evident that the spatial spectra start from $\omega$/c=$k_{oc.}$ and $k_z$ is respectively negative and positive in the spectrum range of $k_{oc.}$ <$\omega$/c <$2k_{oc.}$ and $\omega$/c> $2k_{oc.}$, indicating the SLPTP contains both forward- and backward-propagating parts. Notably, various toroidal pulse solutions, including Ziolkowski's solution [40]-[41], Hellwarth-Nouchi's solution [6],[42]-[48], and Lekner's oscillatory solution [58], of the wave equation contain both forward- and backward-propagating parts, i.e., with both $z$ − c$t$ and $z$ + c$t$ in their waveforms [53]. To illustrate the propagation characteristics of the forward-propagating components of the SLPTP, we isolated spectral components with $\omega$/c> $2k_{oc.}$ and subsequently conducted an inverse transformation to derive the corresponding spatio-temporal toroidal pulses. Remarkably, the SLPTP composed solely of forward-propagating components demonstrates analogous strongly longitudinal polarization traits to those containing both forward and backward-propagating elements, thereby enabling the practical generation of propagation-invariant SLPTPs. For further elucidation, please refer to the supplementary materials.

## Conclusion

In summary, our study has showcased that SLPTPs exhibit both strongly longitudinal polarization and propagation-invariance characteristics. These unique polarization and propagation attributes enable SLPTPs to propagate robustly, manifesting nondiffracting behavior. In contrast to other forms of non-transverse structured light, such as optical knots [1], skyrmions [20], hopfions [12]-[14], Möbius strips [15]-[18], and elementary toroidal pulses [42]-[48], SLPTPs feature a notably pronounced longitudinally polarized component, potentially paving the way for new light-matter interactions. Moreover, unlike strongly longitudinal polarization fields generated through tight focusing [54]-[56], the strongly longitudinal polarization field of SLPTPs can endure upon propagation, thus holding promise for applications in far-field superresolution microscopy. Furthermore, the nondiffracting nature of SLPTPs suggests potential applications in long-distance energy and information transmission. The independent longitudinal polarization component adds a new dimension beyond the dual transverse polarization commonly used in current communication systems, offering prospects for constructing high-capacity wireless communication systems with three independent polarizations.

## Acknowledgements

The authors acknowledge the supports of the the National Natural Science Foundation of China (62171081, 61901086, U2341207), the Aeronautical Science Foundation of China (2023Z062080002), and the Natural Science Foundation of Sichuan Province (2022NSFSC0039). Y. Shen also acknowledges the support from Nanyang Technological University Start Up Grant, Singapore Ministry of Education (MOE) AcRF Tier 1 grant (RG157/23), and MoE AcRF Tier 1 Thematic grant (RT11/23).

## Contributions

## Competing interests

The authors declare no competing interests.

## Data and materials availability

The data that support the findings of this study are available from the corresponding author upon reasonable request.

## Additional information

**Supplementary information** is available for this paper.

Supplementary Information for

# **Progation-invariant strongly longitudinally polarized toroidal pulses**

Ren Wang[1,2*], Ding-Tao Yang[1], Tao Xin[1], Shuai Shi[1], Bing-Zhong Wang[1] and Yijie Shen[3,4*]

[1] Institute of Applied Physics, University of Electronic Science and Technology of China, Chengdu 611731, China

[2] Yangtze Delta Region Institute (Huzhou), University of Electronic Science and Technology of China, Huzhou 313098, China

[3] Division of Physics and Applied Physics, School of Physical and Mathematical Sciences, Nanyang Technological University, Singapore 637378, Singapore

[4] Centre for Disruptive Photonic Technologies, The Photonics Institute, Nanyang Technological University, Singapore 637378, Singapore

*Correspondence to: Ren Wang (rwang@uestc.edu.cn) and Yijie Shen (yijie.shen@ntu.edu.sg)

**Supplementary Note 1: Derivation of the fields of toroidal pulse.**

In 1985, Ziolkowski obtained an exact solution to the scalar wave equation $\{\Delta - \partial_{ct}^2\}\Phi(\boldsymbol{r},t) = 0$ in real space, namely the complex traveling-center-wave basis functions [1]:

$$\Phi(\boldsymbol{r},t) = \frac{\exp[ik\sigma - k\rho^2 / (b+i\tau)]}{4\pi i(b+i\tau)} \tag{1}$$

where $\tau = z - ct$, $\sigma = z + ct$, $\rho^2 = x^2 + y^2$, $k$ is the wavenumber, c is the light speed. An exact electromagnetic directed-energy pulse trains (EDEPT) solution can be superposed by complex traveling-center-wave basis functions [2]:

$$\begin{aligned}
\psi(\boldsymbol{r},t) &= \int_0^\infty \Phi_k(\boldsymbol{r},t)\,\mathrm{F}(k)dk \\
&= \frac{1}{4\pi i(b+i\tau)} \int_0^\infty dk\,\mathrm{F}(k)e^{-ks(\rho,z,t)}
\end{aligned} \tag{2}$$

where $s = \rho^2 / (b+i\tau) - i\sigma$.

In 2004, Lekner derived a new solution by setting $\int_0^\infty dk F(k)e^{-ks} = f(s) = abe^{-k_\infty s} / (s+a)$ [3]:

$$\begin{aligned}
\psi(\boldsymbol{r},t) &= \frac{f(s)}{b+i\sigma} \\
&= \frac{abe^{-k_\infty s}}{(s+a)[b+i\sigma]}\psi_0 \\
&= \frac{abe^{-k_\infty s}}{(\dfrac{\rho^2}{b+i\tau} - i\sigma + a)[b+i\sigma]}\psi_0 \\
&= \frac{abe^{-k_\infty s}}{\rho^2 + [a-i\tau][b+i\sigma]}\psi_0
\end{aligned} \tag{3}$$

where $\psi_0$ represents a constant coefficient.

Setting $q_1 = b$, $q_2 = a$ and $\psi_0 = 1$, a generating function for a general toroidal pulses with a positive oscillatory wavenumber $k_{oc.}$ can be obtained:

$$\psi(\rho, z, t) = \frac{q_2 q_1 e^{ik_{oc.}\sigma - \frac{k_{oc.}\rho^2}{q_1 + i\tau}}}{\rho^2 + [q_2 - i\sigma][q_1 + i\tau]} \qquad (4)$$

where, the parameters $q_2$ and $q_1$ represent respectively the depth of the focal region and effective wavelength when compared to a Gaussian beam. When $k_{oc.} = 0$, Eq. (4) simplifies to the generating function for elementary toroidal pulses [1],[2],[4].

From the scalar wave equation solutions, solutions to Maxwell's equations can be obtained by using Hertz potential $\boldsymbol{A} = \nabla \times [0, 0, \psi] = [\partial_y, \partial_x, 0]\psi$ [2],[5]. The electric and magnetic fields of transverse electric (TE) toroidal pulses are represented by

$$\begin{cases} \boldsymbol{E} = c^{-1}\partial_t \boldsymbol{A} = [-c^{-1}\partial_y\partial_t, c^{-1}\partial_x\partial_t, 0]\psi \\ \boldsymbol{B} = \nabla \times \boldsymbol{A} = [\partial_x\partial_z, \partial_y\partial_z, -\partial_x^2 - \partial_y^2]\psi \end{cases} \qquad (5)$$

Transverse magnetic (TM) toroidal pulses can be obtained through the dual transformation $\boldsymbol{E} \to \boldsymbol{B} = \nabla \times \boldsymbol{A}$, $\boldsymbol{B} \to -\boldsymbol{E} = c^{-1}\partial_t \boldsymbol{A}$ of TE toroidal pulses:

$$\begin{cases} \boldsymbol{E} = \nabla \times \boldsymbol{A} = [\partial_x\partial_z, \partial_y\partial_z, -\partial_x^2 - \partial_y^2]\psi \\ \boldsymbol{B} = c^{-1}\partial_t \boldsymbol{A} = [-c^{-1}\partial_y\partial_t, c^{-1}\partial_x\partial_t, 0]\psi \end{cases} \qquad (6)$$

In cylindrical coordinates, the electric and magnetic fields of TM toroidal pulses can be expressed as:

$$\begin{cases} \boldsymbol{E} = \nabla \times \boldsymbol{A} = (\partial_\rho\partial_z, 0, \partial_z^2 - c^{-2}\partial_t^2)\psi \\ \boldsymbol{B} = c^{-1}\partial_t \boldsymbol{A} = (0, c^{-1}\partial_\rho\partial_t, 0)\psi \end{cases} \qquad (7)$$

$$\begin{cases} E_\rho = \partial_\rho \partial_z \psi \\ E_z = (\partial_z^2 - c^{-2}\partial_t^2)\psi \\ B_\theta = c^{-1}\partial_\rho \partial_t \psi \end{cases} \tag{8}$$

Consequently, substituting Eq. (4) into Eq. (8) yields the expressions for the electric and magnetic fields of TM toroidal pulses.

**Supplementary Note 2: Electric field of the SLPTP.**

The isosurfaces of the longitudinally polarized electric component and spatiotemporal evolution of the elementary toroidal pulse with $k_{oc.}=0$ and the SLPTP with $k_{oc.}=1.463$ $q_1$ are shown in Fig. 1S. Here, $k_{oc.}$ defines the degree of transverse divergence of the longitudinally polarized electric components: as the value of $k_{oc.}$ increases, the pulse envelope of the longitudinal components transforms from a pancake-like shape into a dumbbell-like configuration, accompanied by a significant increase in pulse cycles. During propagation, the SLPTP exhibit maxima in the amplitude of its longitudinal components located on the propagation axis, while the amplitudes of the transverse components vanish at the propagation axis. A comparison with Fig. 2 in the main manuscript reveals that the locations of the maximum lobes in the absolute amplitudes of the SLPTP and the elementary toroidal pulse align with the maxima of their respective longitudinal and transverse components. This provides a visual explanation for the strongly transverse and strongly longitudinal characteristics exhibited by the elementary toroidal pulse and the SLPTP, respectively, in Fig. 2.

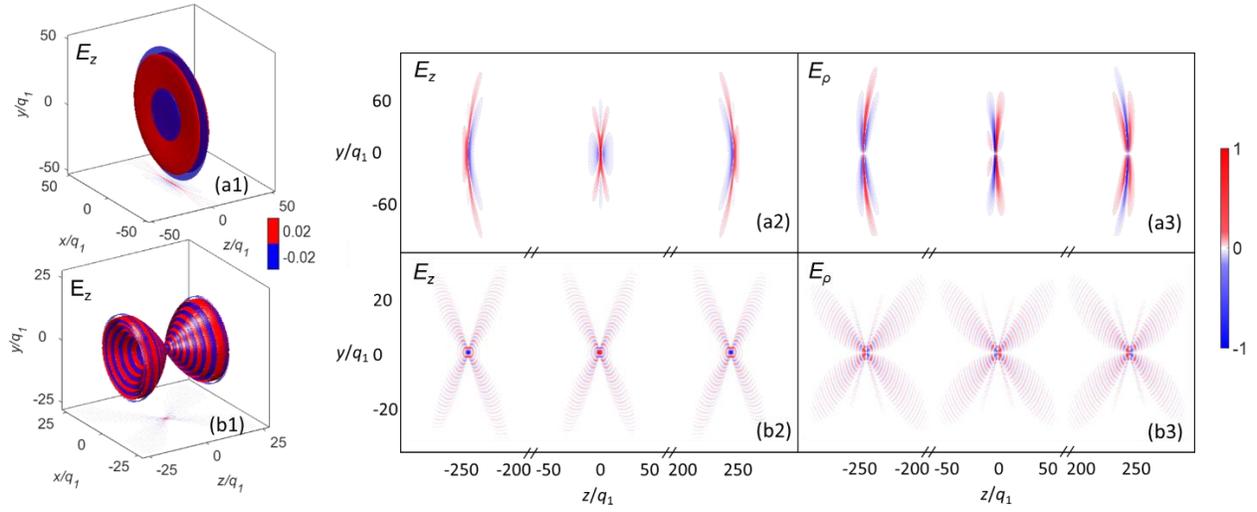

**Fig. 1S: Distribution of longitudinally and transversely polarized electric fields.** (a1~a3) An elementary toroidal pulse with $k_{oc.}=0$ and (b1~b3) a SLPTP with $k_{oc.}=1.463$ $q_1$. (a1 and b1) The isosurfaces with ±0.02 amplitude level and the amplitude distributions of the z-polarized (longitudinally polarized) electric component on x-z plane at $t = 0$. The spatiotemporal evolution of the longitudinally polarized (a2 and b2) and the transversely polarized (a3 and b3) electric components of the pulses on y-z plane.

Fig. 2S presents the transverse and longitudinal components of the electric field as the SLPTP propagates to different positions. It can be observed from the figure that the maxima of these components do not occur simultaneously, exhibiting asynchronous periodic changes in the transverse and longitudinal electric components of SLPTPs. This asynchrony leads to the periodically varying pattern of the ratio $|E_z|_{max}/|E_\rho|_{max}$, as demonstrated in the main manuscript. Despite the asynchronous changes, the amplitude of the longitudinal component remains approximately 11 times larger than that of the transverse component throughout the propagation process, visually illustrating the strongly longitudinal polarization characteristic of SLPTPs.

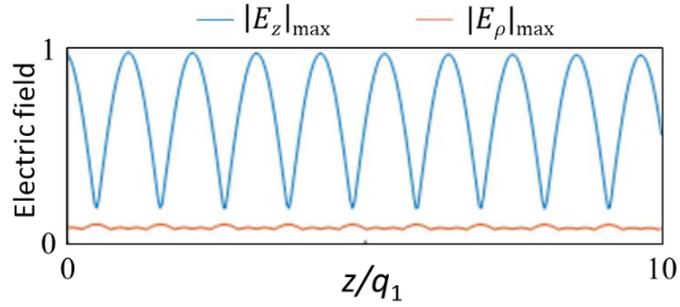

**Fig. 2S  The maxima of transverse and longitudinal components of the electric field as the SLPTP with $k_{oc.}$=1.463 $q_1$ propagates to different positions.** The amplitude of the longitudinal component remains approximately 11 times larger than that of the transverse component throughout the propagation process, although the maxima of transverse and longitudinal components do not occur simultaneously.

**Supplementary Note 3: Magnetic field evolution of the SLPTP.**

The magnetic field $H_\theta$ evolution of the elementary toroidal pulse with $k_{oc.}$=0 and the SLPTP with $k_{oc.}$=1.463 $q_1$ are shown in Fig. 3S. Similar to the electric field, $k_{oc.}$ defines the degree of transverse divergence of the magnetic field: as the value of $k_{oc.}$ increases, the pulse envelope transforms from a pancake-like shape into a dumbbell-like configuration, accompanied by a significant increase in pulse cycles.

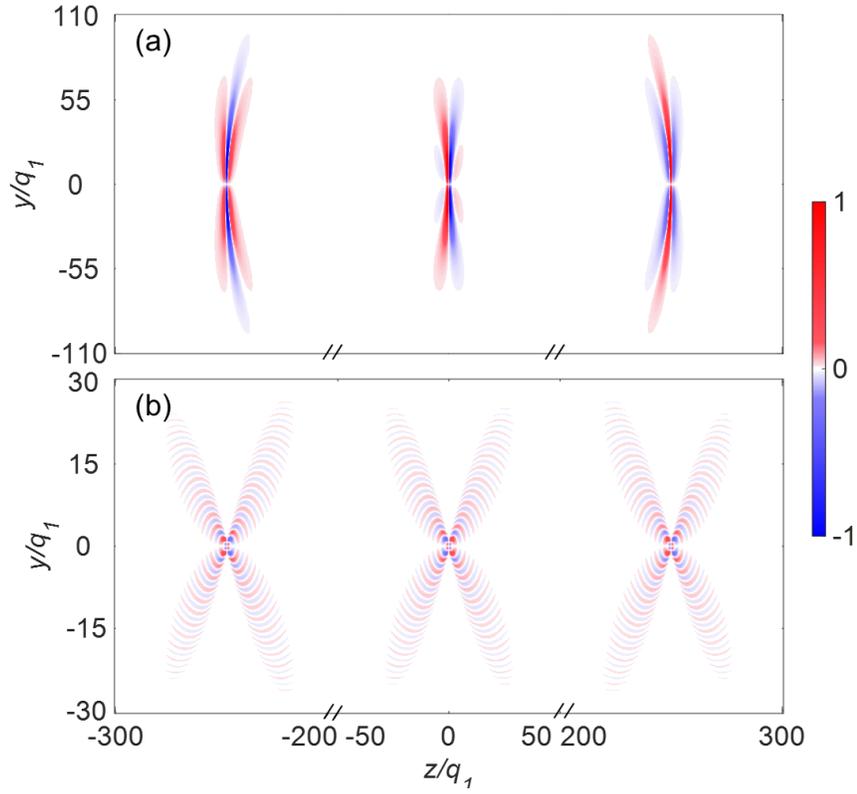

**Fig. 3S  The spatiotemporal evolution of the magnetic field on y-z plane.** (a) An elementary toroidal pulse with $k_{oc.}=0$ and (b) a SLPTP with $k_{oc.}=1.463$ $q_1$. As the value of $k_{oc.}$ increases, the pulse envelope transforms from a pancake-like shape into a dumbbell-like configuration, accompanied by a significant increase in pulse cycles.

**Supplementary Note 4: Frequency spectra of the SLPTP.**

We can perform a Fourier transform on the temporal dimension of the spacetime field $E(\rho,z,t)$ to obtain its frequency spectral distribution:

$$\tilde{E}(\rho,z,\omega)=\int_{-\infty}^{\infty}E(\rho,z,t)\exp[-i(\omega t)]dt \tag{9}$$

The frequency spectra for different $k_{oc.}$ when z=0, 500 $q_1$, and 1000 $q_1$ are depicted in Fig. 4S. At $z$=0, the maximum values of different frequency components of both the elementary toroidal pulse and the SLPTP are situated along the propagation axis ($x$=0). In contrast to the elementary toroidal pulse, the SLPTP exhibits additional sidelobes in its frequency spectrum. Furthermore, the lowest frequency of each spectral distribution is $\omega$=c×$k_{oc.}$. Notably, the spectral distribution of the $E_\rho$ component vanishes at $\omega$=c×2 $k_{oc.}$ for all radii $\rho$, whereas this is not the case for the $E_z$ component. This indicates that the toroidal pulse possesses only a longitudinal field component at $\omega$=c×2 $k_{oc.}$. In addition, the maxima of all spectra of the longitudinal components $E_z$ for all the toroidal pulses with nonzero $k_{oc.}$ are consistently located on the propagation axis $\rho$=0, while those of the elementary toroidal pulse move outwards from the propagation axis. Throughout the propagation process, the frequency spectrum of the SLPTP remains nearly unchanged, demonstrating its propagation-invariant characteristics.

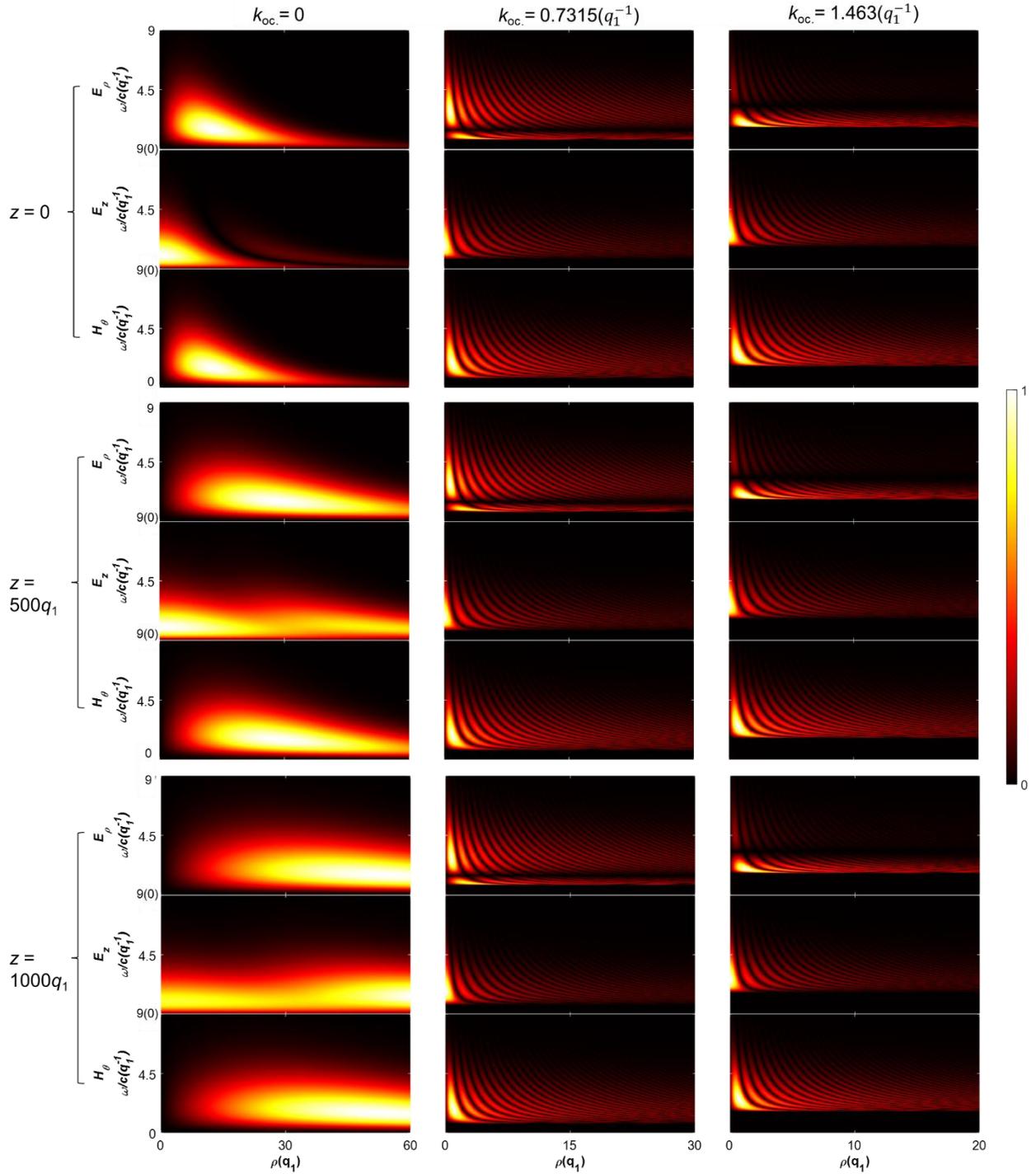

**Fig. 4S** **The frequency spectra for different $k_{oc.}$ when z=0, 500 $q_1$, and 1000 $q_1$.** The maxima of all spectra of the longitudinal components $E_z$ for all the toroidal pulses with nonzero $k_{oc.}$ are consistently located

on the propagation axis $\rho$=0, while those of the elementary toroidal pulse move outwards from the propagation axis.

## Supplementary Note 5: Space-time nonseparability of elementary toroidal pulses and SLPTPs.

We employed the method proposed by Shen et al. to measure the space-time nonseparability of toroidal pulses [6]. This approach leverages the mathematical analogy between classical space-time nonseparability and quantum entanglement. Typically, all temporal characteristics can be fully characterized in the frequency domain; thus, space-time nonseparability can be equivalently interpreted as spatial-spectral nonseparability. The space-time nonseparability of toroidal pulses manifests as isodiffraction properties [7], which indicates that the spatial intensity distribution of each frequency component scales in the same manner along the beam trajectory across all transverse planes perpendicular to the propagation direction.

To illustrate the isodiffraction property, we introduce the spatial normalized intensity of monochromatic light $\lambda_i$ with different wavelengths, denoted as $I(\lambda, r, z) = |E(\lambda, r, z)|^2 / max_{(r, \lambda)} |E(\lambda, r, z)|^2$, where $I(\lambda_i, r_{\lambda i}, z) = max_r I(\lambda_i, r, z)$ represents the peak intensity for each wavelength during propagation. The propagation-invariant characteristic can be observed in the peak intensities $I_m(\lambda, z) = I(\lambda, r_\lambda, z)$ of different wavelength components. To demonstrate the evolution of the transverse profile of the total electric field intensity (integrated over wavelength components), Ref. [6] introduces the normalized radial position $\eta = r/r_{max}(z)$, where $r_{max}(z)$ is the location of the maximum total electric field intensity at a given propagation distance $z$ in the transverse plane. The normalized total intensity curve can be defined as $I_0(\eta, z) = \int I(\lambda, \eta, z) d\lambda / max_r \int I(\lambda, \eta, z) d\lambda$, and both the normalized total intensity and the normalized radial position are independent of the propagation distance. In fact, for isodiffraction pulses, both $I_m(\lambda)$ and $I_0(\eta)$ remain unchanged with propagation distance $z$. Based on these properties, we can introduce two sets of states to describe the space-time nonseparability characteristics within the pulse:

(1) The spectral state, denoted as $|\lambda_i\rangle$ $(i = 1, 2, ..., 20)$, is defined as the monochromatic light state with a wavelength of $\lambda_i$ and a radial position of $r_{\lambda i}$ corresponding to its peak intensity.

(2) The spatial state, designated as $|\eta_i\rangle$, represents the polychromatic light state located at a radial position $\eta = r/r_{max}$, where $r_{max}$ is the radial position at which the total intensity of the optical field reaches its maximum.

Generally, the positions of the spectral and spatial states depend on the propagation distance $z$. For convenience, we can employ the normalized radial position $\eta = r/r_{max}$. Therefore, on the transverse plane at a propagation distance of $z$, the spectral state $|\lambda_i\rangle$ can be represented by the peak intensity $I_m(\lambda_i, z)$ at the position $\eta_{\lambda i}(z) = r_{\lambda i}(z)/r_{max}$. Similarly, the spatial state $|\eta_i\rangle$ can be expressed by the intensity value $I_0(\eta_i r_{max}, z)$ at the normalized radial position $\eta_i$. The positions of the spatial and spectral states, namely $\eta_i$ and $\eta_{\lambda i}(z)$, define the $\eta - z$ plane. For any polychromatic beam, the trajectory of $|\eta_i\rangle$ is always a vertical line on this plane, but $\eta_{\lambda i}(z)$ can follow an arbitrary trajectory. However, for an ideal isodiffraction pulse, both the spectral and spatial states are represented by linear trajectories, reflecting the propagation invariance of the spatial and spectral intensity profiles. Additionally, in this context, we choose the set of states such that the spatial and spectral states coincide exactly, i.e., for an isodiffraction pulse, $\eta_{\lambda i}(z) = \eta_i$.

The spectral state and the spatial state can be represented as

$$\begin{cases} \mathcal{E}_{\lambda_i}(r,z) = \sqrt{I(\lambda_i, r, z)} H\left(r - \delta_{i-1}^{(\lambda)}\right) H\left(\delta_i^{(\lambda)} - r\right) \\ \mathcal{E}_{\eta_i}(r,z) = \sqrt{I_0(r,z)} H\left(r - \delta_{i-1}^{(\eta)}\right) H\left(\delta_i^{(\eta)} - r\right) \end{cases}$$ (10)

where $H(r)$ represents the Heaviside step function, which takes the value of $H(r) = 1$ when the condition $r > 0$ is satisfied, and zero otherwise. The sets of spectral and spatial states are orthogonal, as expressed by $\langle \lambda_i | \lambda_j \rangle = \delta_{ij}$ and $\langle \eta_i | \eta_j \rangle = \delta_{ij}$, where $\delta_{ij}$ denotes the Kronecker delta function. The inner product between the two states is given by $\langle \eta_i | \lambda_j \rangle = \int \mathcal{E}_{\eta i} \mathcal{E}_{\lambda j}^* dr$. In this context, the definition of the classical field $\mathcal{E}_{\lambda i}, \mathcal{E}_{\eta i}$ has already incorporated the radial coordinate $r$.

Next, we introduce the concept of spatially-spectrally inseparable states to quantitatively describe pulses that possess prescribed space-time nonseparability characteristics. The spatial-spectral states can be expressed as

$$|\psi\rangle = \sum_{i=1}^{n} \sum_{j=1}^{n} c_{i,j} |\eta_i\rangle |\lambda_j\rangle \left( c_{i,j}^2 = \langle \eta_i | \lambda_j \rangle, \sum_{ij} c_{ij}^2 = 1 \right)$$ (11)

We employ two metrics to characterize the degree of space-time nonseparability.

(1) Concurrence (con) [8]. In quantum mechanics, con serves as a continuous measure of nonseparability for two-dimensional entangled states. This concept has been generalized to higher dimensions and is defined as $C = \sqrt{2\left[1 - Tr\left(\rho_A^2\right)\right]}$, where $\rho_A$ represents the density-reduced matrix [9]. For any $d$-dimensional state, the normalized con $C/v_d$, where $v_d = \sqrt{2(1 - 1/d)}$, ranges from 0 to 1, indicating the absence of entanglement (or complete separability) and strong nonseparability (maximum entanglement).

(2) Entanglement of formation (EoF) [10]. In quantum mechanics, EoF is another common metric used to quantify quantum entanglement. It is computed from the reduced density matrix $E = -Tr\left(\rho_A \log_2\left(\rho_A\right)\right)$, and is typically normalized to $E / \log_2\left(d\right)$ in the $d$-dimensional case.

To compare the diffraction characteristics of the elementary toroidal pulse and the SLPTP during propagation, Fig. 5S compares the traces of the radial positions where the spectral intensity of the transversely polarized component reaches its maximum for different frequencies and the η-z map of spectral and spatial states of toroidal pulses with $k_{oc.}$=0 and $k_{oc.}$=1.463 /$q_1$. The spectral components of the elementary toroidal pulse undergo a focusing process followed by diffusion during propagation. In contrast, the frequency spectral components of the SLPTP exhibit no significant diffusion effect throughout the entire propagation process and remain nearly parallel, indicating the quasi-nondiffractive nature of the SLPTP. While the elementary toroidal pulse and the SLPTP possess distinct diffraction speeds, the η-z maps of spectral and spatial states of the two kinds of pulses show isodiffractive behavior [6], as also shown in Table 1S.

In Table 1S, we summarize the entanglement calculation results. It can be observed that both the the elementary toroidal pulse with $k_{oc.}$=0 and the SLPTP with $k_{oc.}$=1.463 $q_1$ exhibit remarkably high con and EoF, indicating that it is in a state of near-maximum entanglement. The slight deviations of the entanglement measurements from unity are attributed to different intensity levels across various spatial and spectral states. While the elementary toroidal pulse and the SLPTP possess distinct diffraction speeds, they both exhibit isodiffractive behavior.

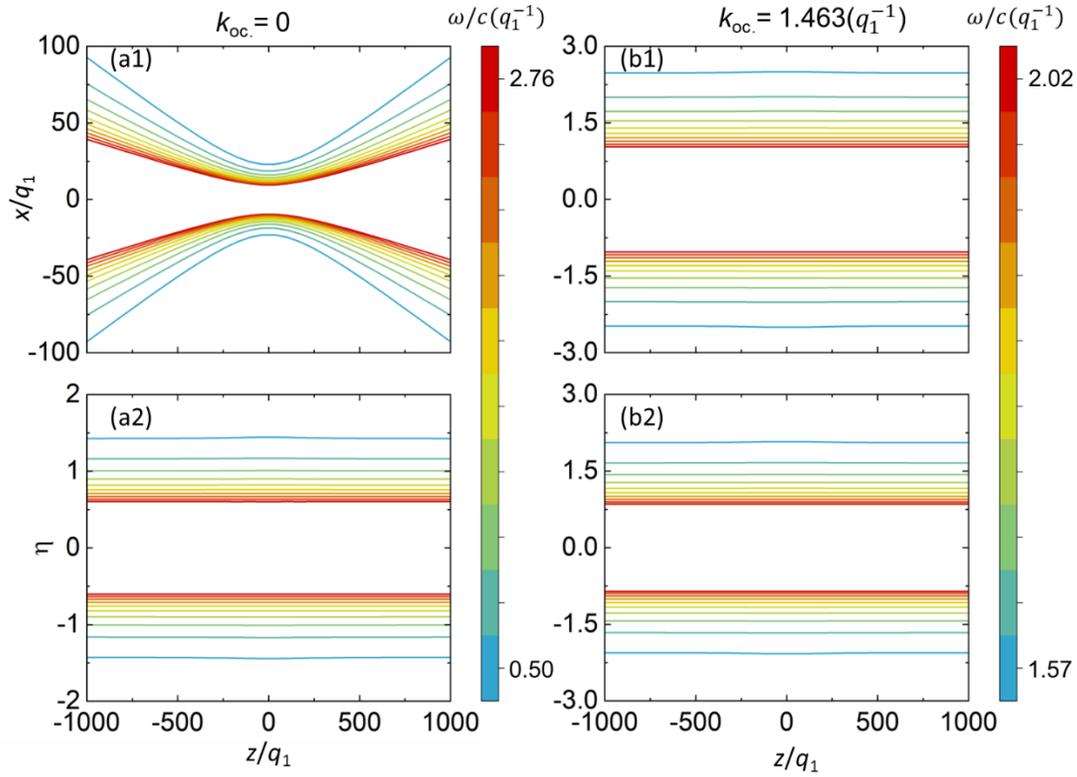

**Fig. 5S  Spatial states of toroidal pulses.** (a1 and b1) The traces of the radial positions where the spectral intensity of the transversely polarized component reaches its maximum for different frequencies and (a2 and b2) the η-z map of spectral and spatial states of the toroidal pulses with (a1 and a2) $k_{oc.}$=0 and (b1 and b2) $k_{oc.}$=1.463 /$q_1$.

**Table 1S  Parameter comparison of toroidal pulses with different resonant wavenumber.** N-Con and N. EoF represent con and EoF in intensity-normalized measurement, respectively.

| $k_{oc.}$ / $q_1^{-1}$ | **0** | **1.463** |
|---|---|---|
| **N-Con** | 0.9997 | 0.9864 |
| **N-EoF** | 0.9988 | 0.9494 |

**Supplementary Note 6: Plane wave expansion method.**

In the realm of electromagnetic wave analysis, the three-dimensional plane wave spectra of $\tilde{E}(k_\rho, k_z, \omega)$ can be obtained through three-dimensional Fourier transformation.

$$\tilde{E}(k_\rho, k_z, \omega) = \int_{-\infty}^{\infty} \int_{-\infty}^{\infty} \int_{-\infty}^{\infty} E(\rho, z, t) \exp[-i(k_r \rho + k_z z + \omega t)] d\rho dz dt \qquad (12)$$

Subsequently, employing projections facilitates the derivation of corresponding two-dimensional plane wave spectra from the aforementioned three-dimensional spectra.

In order to illustrate the propagation attributes specific to the forward-propagating components of the SLPTP, we nullify the portions of the field within the plane wave spectra where $k_z$ is less than zero. This manipulation yields $\tilde{E}'(k_\rho, k_z, \omega)$.

$$\tilde{E}'(k_\rho, k_z, \omega) = \begin{cases} \tilde{E}, k_z > 0 \\ 0, k_z \leq 0 \end{cases} \qquad (13)$$

Subsequent to this adjustment, an inverse Fourier transformation is performed on $\tilde{E}'(k_\rho, k_z, \omega)$ to procure a novel spatiotemporal field denoted as $E'(r, z, t)$. A comparative analysis between the original spatiotemporal field $E(r, z, t)$ and its modified counterpart $E'(r, z, t)$ enables an examination of the characteristics inherent in the SLPTP with exclusively forward-propagating components juxtaposed against those featuring both forward and backward-propagating components. The comparative analysis reveals that the SLPTP with solely forward-propagating components demonstrates analogous strongly longitudinal polarization characteristics to its counterpart containing both forward and backward-propagating components.

**Supplementary Note 7: Forward-propagating characteristics of the SLPTP.**

To demonstrate the propagation characteristics of the forward-propagating parts of the SLPTP, we extracted the spectral components with $\omega/c > 2k_{oc.}$ and subsequently performed an inverse transformation to obtain the spatio-temporal toroidal pulse field. Fig. 6S presents the propagation behavior of the SLPTP solely containing forward-propagating components, carring over 80% energy of the total pulse. In comparison to Fig. 3, it can be observed that the SLPTP with only forward-propagating parts exhibits similar strongly longitudinal polarization characteristics to the SLPTP containing both forward and backward-propagating parts. Specifically, when $k_{oc.}$ lies within the range of $1/q_1 \sim 2/q_1$, the longitudinal component of the toroidal pulses significantly dominates the transverse component throughout the propagation process.

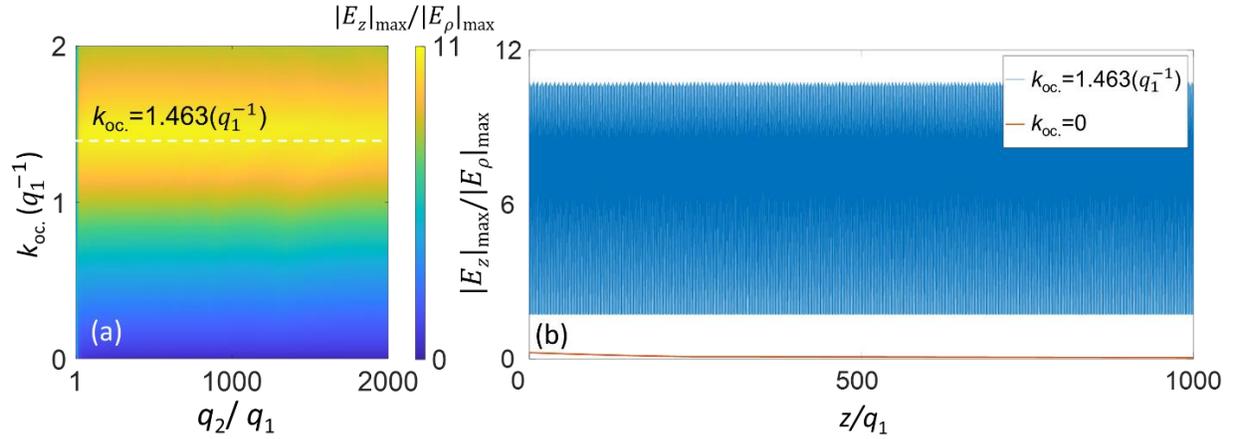

**Fig. 6S The ratio of the maximum values of longitudinal and transverse electric components $|E_z|_{max}$ and $|E_\rho|_{max}$ of the toroidal pulsing only containing forward- propagating parts with different oscillatory wavenumber $k_{oc.}$.** (a) Versus $q_2/q_1$ and (b) versus the pulse center position at different $z$-values. For (a), z=0; for (b), $q_2=500q_1$.

**Supplementary Video**

Supplementary Video 1: Electric field vector evolution of elementary toroidal pulses.

Supplementary Video 2: Electric field vector evolution of SLPTPs.

Supplementary Video 3: Propagation of the transverse polarized electric field $E_\rho$ in yz plane of elementary toroidal pulses.

Supplementary Video 4: Propagation of the transverse polarized electric field $E_\rho$ in yz plane of SLPTPs.

Supplementary Video 5: Propagation of the longitudinal polarized electric field $E_z$ in yz plane of elementary toroidal pulses.

Supplementary Video 6: Propagation of the longitudinal polarized electric field $E_z$ in yz plane of SLPTPs.

Supplementary Video 7: Propagation of the 3D transverse polarized electric field $E_\rho$ of elementary toroidal pulses.

Supplementary Video 8: Propagation of the 3D transverse polarized electric field $E_\rho$ of SLPTPs.

Supplementary Video 9: Propagation of the 3D longitudinal polarized electric field $E_z$ of elementary toroidal pulses.

Supplementary Video 10: Propagation of the 3D longitudinal polarized electric field $E_z$ of SLPTPs.